# AWRP: Adaptive Weight Ranking Policy for Improving Cache Performance

Debabala Swain, Bijay Paikaray and Debabrata Swain

**Abstract—** Due to the huge difference in performance between the computer memory and processor , the virtual memory management plays a vital role in system performance .A Cache memory is the fast memory which is used to compensate the speed difference between the memory and processor. This paper gives an adaptive replacement policy over the traditional policy which has low overhead, better performance and is easy to implement. Simulations show that our algorithm performs better than Least-Recently-Used (LRU), First-In-First-Out (FIFO) and Clock with Adaptive Replacement (CAR).

**Index Terms—** Adaptive page replacement, CAR, FIFO, LRU, Memory management, Weight ranking.

—————————— ◆ ——————————

## 1 INTRODUCTION

In modern computer systems the efficiency of the memory hierarchy is one of the most crucial issues. To obtain a high-performance memory system, caching is the best direction in terms of cost and effectiveness. The performance of the cache memories to compensate the speed gap between processors and main memory is determined by the access time and hit rates. Caching is an old and well-known performance enhancement technique that is widely used in computer systems. Processor's cache, paged virtual memory and web caching are some instances of using this technique. In all levels of storage hierarchy, performance of the system is related to cache mechanism. The time access for cache is at least 10 times smaller than the main memory.

Increasing the cache size results in better performance, but resulting a very expensive technology. That's why the size of cache memory is always smaller than the main memory in every system. According to cache memory access time, fetching data from cache is a big advantage and has a significant role in system performance. We must use other techniques to make cache more efficient for systems which make use of it in their memory hierarchy

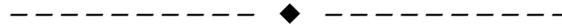

- *Debabala Swain is with CSE Dept, Centurion University of Technology and Management, Bhubaneswar Campus, Orissa, India.*

- *Debabrata Swain is with Dept of Computer Science, Berhampur University, Orissa, India.*
- *Bijay k Paikaray is with CSE Dept, Centurion University of Technology and Management, Bhubaneswar Campus, Orissa, India.*

The terminology used for cache efficiency is "Hit Rate" (Hit Rate - represents the rapport between the number of addresses accessed from cache and the total number of addresses accessed during that time). The antonym for "Hit Rate" is named "Miss Rate" which can be determined by Miss Rate = 1 – Hit Rate formulas. [1,24]

Cache efficiency can be changed by implementing different replacement policies. Cache and auxiliary memories use uniformly sized items which are called pages. When the cache is empty the first access will be always a miss. The misses may occur until the cache is filled with pages. After some cycles, when the cache is filled, this kind of misses will not occur anymore. Another kind of miss occurs when the cache is full of pages and there is an access to a page that is not in the cache, in this case replacement policy should specify conditions under which a new page will be replaced with an existing one. The main role of replacement policy is to choose a page which has the lowest probability of further usage in the cache to be replaced with the new one. Generally there are three ways of mapping memory into cache: 1) Direct Mapped in which each block from the main memory is mapped only to a unique cache block, no matter whether that block is empty or not. This model of cache organization doesn't require a replacement policy. 2) Fully Associative in which each block from the main memory can be mapped to any of cache blocks. If the cache is full then a replacement policy needs to decide which data will be invalidated from cache for making room of this new data block. 3) Set Associative in which the cache is split into many sets. The new data can be mapped only to the blocks of a certain set, determined by some bits from address field. In practice it is used mostly because of a very efficient ratio between implementation and efficiency. There may be another kind of miss in the direct map caches which is due to the same data found in many places in the cache. [1-9]

An optimal replacement policy must have good performance with low overhead on the system and it must be easy to implement in hardware. It means, there is a trade-off between a replacement policy hit ratio, its implementation difficulties and its overhead cost.

The basic idea of our algorithm is to *weight and rank pages based on their recency index, frequency index.* So, pages that are more recent and have been used frequently are ranked at the higher level. Pages with small frequency but better recency will rank higher than the pages with lower recency as well as low frequency. It means that the probability of replacing pages with smaller weight is more than the one with higher weight.

The best advantage of our algorithm is that it can behave like both LRU and LFU by replacing pages that were not recently used and pages that are accessed very few. Like LRU we don't need to find out the recency factor $R_i$ for each object i, which needs to obtain a i x i matrix for a cache of block size i. It is a time and space consuming method.

Our policy can perform better than LRU, FIFO and CAR. ***We call this algorithm policy as Adaptive-Weight-Ranking-Policy (AWRP), which means the algorithm, will assign weight to each object in the buffer and rank each object as per the weight***.

This paper is organized as follows:
In Section 2, we describe a comparative study of previous works in page replacement policies.
In Section 3, we propose our method and its details. In Section 4, we show implementation and simulation results with performance analysis to prove the advantages of our replacement policy. At the end, we discuss about future work in section 5.

## 2 BACKGROUND

Caching is a mature technique that has been widely applied in many computer science areas, Operating Systems and Databases are two most important ones. Currently, the World Wide Web is becoming another popular application area of caching. The LRU Algorithm has been widely used in commercial systems as page replacement policy and it could be the first choice for those systems. It is based on the assumption that pages which has not been referenced for the longest time won't be used in near future. In other word, the reference probability of pages that have been heavily used is more than the other pages.

*Under the* **LRU** (Least Recently Used) algorithm, is based on the observation that pages that have been heavily used in the last few instructions will probably be heavily used again in the next few. Conversely, pages that have not been used for ages will probably remain unused for a long time. This idea suggests a realizable algorithm: when a page fault occurs, throw out the page that has been unused for the longest time. This strategy is called **LRU** (**Least Recently Used**) paging.

The algorithm LRU has many disadvantages [11]:

1. On every hit to a cache page it must be moved to the most recently used (MRU) position. In an asynchronous computing environment where multiple threads may be trying to move pages to the MRU position, the MRU position is protected by a lock to ensure consistency and correctness. This lock typically leads to a great amount of contention, since all cache hits are serialized behind this lock. Such contention is often unacceptable in high performance and high throughput environments such as virtual memory, databases, file systems, and storage controllers.

2. In a virtual memory setting, the overhead of moving a page to the MRU position–on every page hit–is unacceptable [24].

3. While LRU captures the "recency" features of a workload, it does not capture and exploit the "frequency" features of a workload [25, p. 282]. More generally, if some pages are often requested, but the temporal distance between consecutive requests is larger than the cache size, then LRU cannot take advantage of such pages with "long-term utility".

4. LRU can be easily polluted by a scan, that is, by a sequence of one-time use only page requests leading to lower performance.

In **FIFO** (First-In-First-Out) the operating system maintains a list of all pages currently in memory, with the page at the head of the list the oldest one and the page at the tail the most recent arrival. On a page fault, the page at the head is removed and the new page added to the tail of the list may be a good example which has very simple implementation, but gets into problem when the size of physical memory is big. The problem with FIFO is that it ignores the usage pattern of the program [1].

**LFU** (Least Frequently Used) is a frequency-based policy, in which the page with low frequency will be replaced first. It works badly because different parts of memory have different time–variant patterns.

The LFU policy has several drawbacks: it requires logarithmic implementation complexity in cache size, pays almost no attention to recent history, and does not adapt well to changing access patterns since it accumulates stale pages with high frequency counts that may no longer be useful [1-8].

*ARC* (Adaptive Replacement Cache) combines the LRU and LFU solutions and dynamically adjusts between them. ARC like LRU is easy to implement and has low over-head on systems[9]. LRU captures only recency and not frequency, and can be easily "polluted" by a scan. A scan is a sequence of one-time use only page requests, and leads to lower performance.

ARC overcomes these two downfalls by using four doubly linked lists. Lists T1 and T2 are what is actually in the cache at any given time, while B1 and B2 act as a second level. B1 and B2 contain the pages that have been thrown out of T1 and T2 respectively. The total number of pages therefore needed to implement these lists is 2*C*, where *C* is the number of pages in the cache. T2 and B2 contain only pages that have been used more than once. The lists both use LRU replacement, in which the page removed from T2 is placed into B2. T1 and B1 work similarly together, except where there is a hit in T1 or B1 the page is moved to T2. The part that makes this policy very adaptive is the sizes of the lists change. List T1 size and T2 size always add up to the total number of pages in the cache. However, if there is a hit in B1, also known as a Phantom Hit (i.e. not real hit in cache), it increases the size of T1 by one and decreases the size of T2 by one. In the other direction, a hit in B2 (Phantom Hit) will increase the size of T2 by one and decrease the size of T1 by one. This allows the cache to adapt to have either more recency or more frequency depending on the workload [9].

ARC has following two drawbacks:

1) The overhead of moving a page to the most recently used position on every page hit.

2) Serialized, in the fact that when moving these pages to the most recently used position the cache is locked so that it doesn't change during this transition.

The **CAR** (Clock with Adaptive Replacement) combines the LRU and LFU solutions and dynamically adjusts between them. ARC like LRU is easy to implement and has low over-head on system.

CAR only overcomes the first downfall, the second still has a presence. CAR uses two clocks instead of lists, a clock is a circular buffer. The two circular buffers are similar in nature to T1 and T2, and it contains a B1 and B2 as well. The main difference is that each page in T1 and T2 contains a reference bit. When a hit occurs this reference bit is set on. T1 and T2 still vary in size the same way they do in ARC (i.e. Phantom Hits cause these changes). When the cache is full the corresponding clock begins reading around itself (i.e. the buffer) and replaces the first page whose reference bit is zero. It also sets the reference bits back to zero if they are currently set to one. The clock will continue to travel around until it finds a page with reference bit equal to zero to replace [11].

## 3 WEIGHT RANKING ALGORITHM

In this section we will use the standard replacement policy terminologies. We assume that the size of the blocks is equal and the replacement algorithm is used to manage a memory that holds a finite number of these blocks. A *hit* will occur when there is a reference to a block in the buffer. When we have a reference to a block not in the buffer, a *miss* will occur and referenced block must be added to buffer. When buffer is full and a miss occurs, an existing block must be replaced to make room for a new one.

For ranking of the referenced blocks, we consider three factors. Let $F_i$ be the frequency index which holds the frequency of each block *i* in the buffer and $R_i$ be the recency index which shows the clock of last access when buffer has been referenced and $N$ is the total number of access to be made. Then the weighting value of block *i* can be computed as:

$$W_i = \frac{F_i}{N - R_i} \quad (1)$$

Initially all the indices of $R_i$, $F_i$, $W_i$ are set to **0**.

*1) Whenever a page frame k is referenced, the hardware increments corresponding $F_i$ by 1. When a hit occurs the respective $R_i$ counter will hold the corresponding clock access value. At any instant, the block whose weight index is lowest is the least recently used block, the block with next higher weight index is the next least recently used, and so forth.*

*2) When a miss occurs i.e. a new object k is placed in the i block then $F_i$ and $R_i$ for the new block will be updated. Because, it means that object k has been placed and once referenced.*

In every access to buffer, if referenced block *i* is in the buffer then a ***hit*** is occurred and our policy will work in this way:

1) $F_i$ will be changed to $F_i + 1$ for every *i* and $R_i$ will hold the corresponding clock access value.

But if referenced block *i* is not in the buffer then a ***miss*** occurs and our policy will choose the block in buffer, by calculating the weight function i.e. searching for object with smallest weighting value. It will be evaluated as follows:

1) The $W_i$ value has to be evaluated for each block *i*, for every $N \neq R_i$.

2) Now the page *i* with lowest weight index $W_i$ will be replaced with *k*.

Then the corresponding $R_k$ will be the clock value, $F_k = 1$ and $W_k$ will be set to **0**.

*The weighting value of blocks that are in buffer will only be updated when a miss occurs in the buffer because one of the important concepts in replacement policies is its overhead in the systems.* When cache size is small our algorithm will all most work like **LRU**. But in larger cache size it works better than LRU, FIFO and CAR.

We have only talked about how our replacement algorithm works, but we have not discussed how it can be implemented in real system.

AWRP needs three elements to work and will add space overhead to system: first, algorithm needs a space for counter $F_i$ second, it needs a space for counter $R_i$ and third it needs a counter for $W_i$, which holds weighting value for each object in the buffer. Calculating weighting function value for each object after every access to cache will cause a time overhead to system. We have to consider the $W_i$ as floating numbers.

## 4 PERFORMANCE ANALYSIS

To evaluate our replacement algorithm experimentally, we simulated our policy and compared it with other policies like LRU, FIFO and CAR. The simulator program was designed to run some traces of load instructions executed for some real programs and implement different replacement policies with different cache sizes. The obtained *hit ratio* depends on the replacement algorithm, cache size and the locality of reference for cache requests. Modular design of simulator program allows easy simulation and optimization of the new algorithm.

### 4.1 Input traces

Our address trace is simply a list of one thousand memory addresses produced by a real program running on a processor. Generally address traces would include both instruction fetch addresses and data (load and store) addresses, but we are simulating only a data cache, so these traces only have data addresses. The mapping scheme is considered and set to *set associative* mapping. According to traces that have been used, we considered 8 cache sizes and ran the simulation program to test the performance of our proposed algorithm.

### 4.2 Simulation Results

We executed simulation program for all 8 different cache sizes and compared it with LRU, FIFO, and CAR. As it is indicated in Figure1, our Adaptive weighting ranking policy performs clearly better than LRU, FIFO and CAR. The average enhancement compare with **LRU** is about **1.22**%, average enhancement for **FIFO** is about **0.95**% and average enhancement for **CAR** is about **0.47**%.In Table 1, we can see the accurate hit ratio for all page references for all four schemes.

For the other three cases we executed simulator program with same 8 cache sizes. The result shows that the maximum gain of hit ratio of **17.47**% from LRU, is attained for the cache size of 90 from LRU, and minimum gain is **0.76**% for cache size of 30 from LRU case. The result shows that the maximum hit ratio for AWRP is about **75.42**% for a cache size of 210, and minimum hit ratio is **41.9**% for cache size of 30.

If we will compare AWRP with individual algorithm then maximum gain from LRU is **17.47** % at cache size 90 and minimum gain is **0.77**% at cache size 30, maximum gain from FIFO is **11.47**% at cache size 120 and minimum gain is **1.88**% at cache size of 210.Maximum gain from CAR is **8.75**% at cache size 60 and minimum gain is **0.96**% at cache size of 150.One exceptional case is at a cache size of 180 CAR works **0.93**% better than AWRP. Also at cache size of 210 CAR and AWRP give same hit ratio of **75.42**%.

In the used trace the hit ratio of proposed algorithm is always better than LRU, FIFO and CAR replacement policies, and in the worse cases the result is equal to the maximum hit ratio of CAR. That is because of the considered weighting factors in AWRP which never let the result become worse than maximum hit ratio of these replacement policies except the 210 cache size from CAR.

| FRAME SIZE(in blocks) | LRU | FIFO | CAR | AWRP |
|---|---|---|---|---|
| 30 | 41.6 | 40.93 | 40.24 | 41.92 |
| 60 | 48.6 | 49.26 | 49.65 | 54.41 |
| 90 | 54.5 | 57.48 | 59.27 | 64.02 |
| 120 | 60.81 | 62.14 | 66.2 | 69.27 |
| 150 | 65.21 | 66.3 | 70.96 | 71.65 |
| 180 | 72.3 | 72.84 | 75.22 | 74.53 |
| 210 | 72.7 | 74.03 | 75.42 | 75.42 |

**Table 1**. A Comparison between hit ratio of LRU, FIFO, CAR and AWRP for 8 different frame sizes.

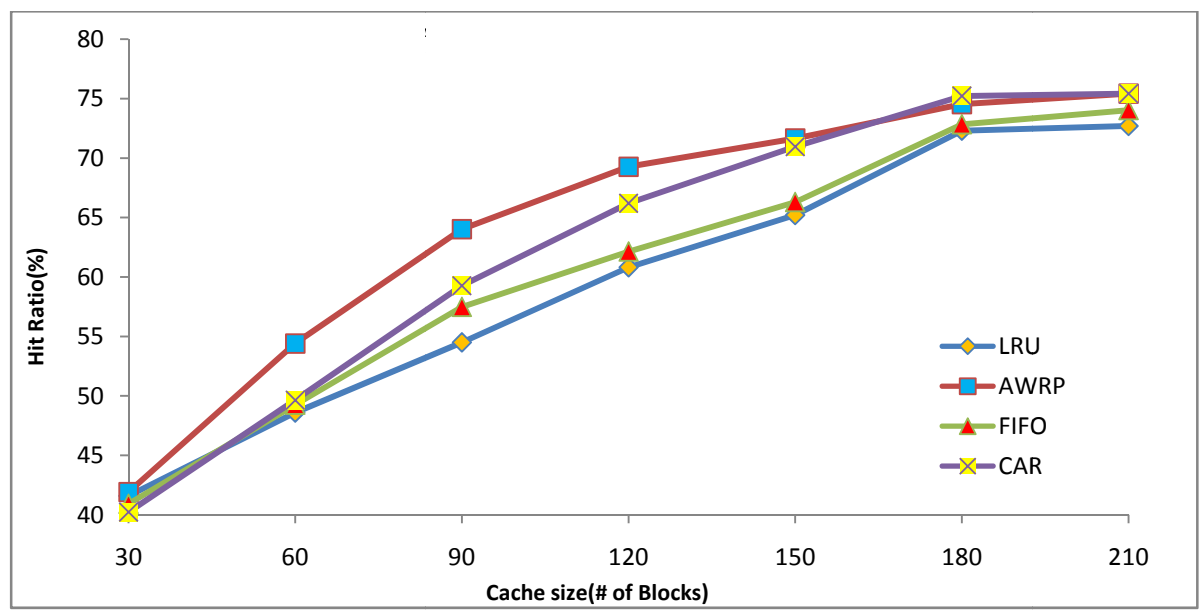

**Figure1**. Performance Analysis of LRU, FIFO, CAR and AWRP with different frame sizes.

## 5 SUMMARY & CONCLUSIONS

In this paper we have introduced a new page replacement policy and the simulation results shows that it is a modification and improvement of famous replacement policies like LRU, FIFO and CAR. We attended to reference rate of each object in the buffer and claimed that the probability of re-accessing of the pages with higher weighting value is more than the others and simulations confirmed our claim. We simulated AWRP and compared it with only three famous policies. This algorithm can be simulated with other policies with benchmark traces and for different applications and then the overall results may be compared with the results of recently proposed replacement policies.

It can be concluded that if the additional parameters and factors which describe features of objects in the buffer be taken into account, then AWRP can be suitably used for applications like database servers, web caching and other applications.

**Authors**

**Debabala Swain,** received her BE in Electronics & Telecommunication Engineering from SMIT, Berhampur, India, in June 2005, and M.Tech in Computer Science from Berhampur University , India in June 2008.Presently pursuing her PhD in Computer Science under Utkal University, Bhubaneswar, India. Currently she is working as an Asst. Professor in Department Computer science at Centurion University of Technology and Management, Bhubaneswar, India .Her prime research interest includes signal processing, wireless networking, System architecture and performance evaluation.

**Debabrata Swain** received his BTech in Computer Science Engg from RIT, Berhampur, India in June 2008.Presently pursuing his M.Tech in Computer Science under Berhampur University**.** His subject of interest includes Computer Architecture, Computer Networking.

**Bijay Paikaray** received his MCA from Sambalpur University in June 2010.Presently working as a Lab Instructor in the Dept. of CSE in Centurion University of Technology and Management, Bhubaneswar, India. His subject of interest includes Computer Organization, Network Security.